\newcommand{\bi}{\bibitem}
\newcommand{\beq}{\begin{equation}}
\newcommand{\eeq}{\end{equation}}
\newcommand{\bea}{\begin{eqnarray}}
\newcommand{\eea}{\end{eqnarray}}
\newcommand{\oncite}{\onlinecite}
\newcommand{\rd}{\rho_{\rm d}}
\newcommand{\tmin}{T_{{\rm min}}}
\documentstyle[aps,prl,multicol,graphicx]{revtex}
\begin{document}
\draft
\preprint{draft}
\title{Weak-localization type description of conduction in the \lq\lq anomalous\rq\rq\/ metallic
state}
\author{B.\ L.\ Altshuler$^{a,b}$, G.\ W.\ Martin$^c$, D.\ L.\
Maslov$^c$, V.\ M.\ Pudalov$^d$, A.\ Prinz$^e$,  G.\ Brunthaler$^e$, G.\
Bauer$^e$}
\address{$^{a)}$NEC Research Institute, 4 Independence Way, Princeton, NJ
08540}
\address{$^{b)}$Physics Department, Princeton University,
Princeton, NJ 08544}
\address{$^{c)}$Department of Physics, University of
Florida\\ P.\ O.\ Box 118440, Gainesville, Florida 32611-8440}
\address{$^{d)}$ P.\ N.\ Lebedev Physics
Institute, 117924 Moscow, Russia.}
\address{$^e$Institut f\"{u}r Hableiterphysik,
Johannes Kepler Universit\"{a}t Linz, A-4040 Austria}

\date{\today}
\maketitle

\begin{abstract}
This paper is devoted to the temperature dependence
of the resistivity in high mobility Si-MOSFET
samples over the wide range of densities in the ``metallic phase'' ($n > n_c$)
but not too close to the critical density $n_c$. Three domains
of qualitatively different behavior in $\rho(T)$ are identified.
These are:
[i] ``quantum domain'' ($T<T_q$), where a logarithmic
temperature dependence of the resistivity (with $d\rho /dT <
0$) dominates;  [ii]
``semi-classical domain''  ($T_{\rm cros}<T<E_{F}$), in which a
a strong variation of  Drude resistivity $\rho(T)$
is observed (with $d\rho/dT>0$); and [ii] ``crossover domain''
between the former two ($T_q<T< T_{\rm cros}$), where
approximately linear temperature dependence dominates
(with $d\rho/dT>0$).
For
high mobility Si-MOS samples we find empirically $T_q \sim 0.007
E_F$ and $T_{\rm cros} \sim 0.07E_F$.
In the crossover regime and at higher
densities  ($\gtrsim 20\times 10^{11}$cm$^{-2}$), $\rho(T)$
goes through a minimum at temperature $\tmin$.
 Both the absolute value of $\tmin$
and its dependence on the carrier concentration
are found to be in a reasonable agreement with the
conventional weak-localization theory. For densities
smaller than $\sim 20\times 10^{11}$cm$^{-2}$, the theoretical
estimate for $\tmin$
falls outside the experimentally accessible temperature range.
This explains the
absence of the minimum at these densities in the data.
In total,  over the two decades in the temperature (domains
[ii] and [iii]), the two semiclassical effects mentioned above
 mimic  the
metallic like transport properties. Our analysis shows that the
behaviour of resistivity $\rho(T)$
in the region of $\rho \ll
h/e^2$ can be described  phenomenologically in terms of the
conventional weak-localization theory.
\end{abstract}
%\newpage
\begin{multicols}{2}
\section{Introduction}
\vspace{-0.2in}
The striking metallic-like (with $d\rho/dT>0$) temperature dependence of the
resistivity observed by now in many two-dimensional (2D)
electron and hole systems
\cite{RvsT} remains a subject of active interest.
The major question  to be addressed in this area is
whether this effect manifests the existence of a metallic ground state
(contrary to the conventional localization theory \cite{gang4}
which predicts that $\rho \rightarrow \infty$ for $T\to 0$ in 2D)
or it is a finite-temperature effect (not necessarily of
a single-particle origin).

The analysis and interpretation of the transport experiments
on  various 2D systems
are  complicated by the
fact that the measurements are performed
at finite though low temperatures. It is shown in this paper
that the rapid metallic-like $\rho(T)$-dependence
is not associated with quantum-interference effects.
The origin of the $\rho(T)$-dependence is thus a semi-classical
effect (in a sense that electron transport can be described by the
classical Boltzmann equation whereas electron statistics may be
degenerate or non-degenerate and individual scattering events may
or may not require quantum-mechanical description). The
microscopic nature of this semi-classical effect is not known at
the moment and it is not a subject of the present paper. (Although
a number of scenarios have been proposed \cite{am,das,meir,yaish},
it is not clear which one or which combination of the above is
capable of explaining all the experimental findings.) We also
demonstrate that the semi-classical $T$-dependence is present down
to to temperatures as low as 100 times less than the Fermi energy
$E_F$, whereas the question on the nature of the ground state is
obviously related to the \lq\lq quantum\rq\rq\/ regime of even
lower temperatures, where all semi-classical effects freeze out.
At finite temperatures, the semi-classical $T$-dependent
scattering in high mobility samples {\em may}  (and {\em does})
{\em mimic metallic-like transport behavior}. This rather complex
and confusing picture explains why numerous controversial
conclusions were reported based on different experimental data.
This paper is concerned with the region of the `metallic' state
not too close to the apparent metal-insulator transition, in which
$\rho \ll h/e^2$. The
behaviour of the 2D carrier system in the
critical region in the vicinity of this transition was considered
earlier by Altshuler et al. \cite{akk} and by Abrahams et al.
\cite{aks2000} from different points of view.

We present here a ``phase diagram'' in
variables of temperature,
density and resistivity
and classify the domains of densities and
temperatures where quantum-interference or semi-classical effects
of different origin %are dominant
manifest themselves.
We analyze the transport data {\em over a wide range of
temperatures and densities}
and show that different semi-classic temperature dependences
dominate in different
domains, down to temperatures as low as $\sim 0.01E_F$,
and mimic ``metallic-like''
conduction.
In particular, we analyze in detail the crossover from
semi-classical to quantum-coherent conduction domains and show that
the data can be described reasonably well with in a
phenomenological model \cite{amp}, based
on the conventional
weak-localization theory.
The present analysis deals mainly with the data obtained on
Si-MOSFET samples in which the metallic conduction is most
strongly pronounced and in which disorder effects can be  revealed by
comparing the samples produced by the same method
but with different  mobilities.
However, we believe that our conclusions may
have a wider applicability, and remain valid, at least
qualitatively, for other material systems.
\vspace{-0.2in}
\section{Identification of the quantum, semi-classical and classical domains}
\subsection{Parameter space}
\vspace{-0.2in}
Figure 1 shows a typical temperature dependence of the resistivity,
$\rho(T)$, measured over a wide range of densities on a high mobility
Si-MOSFET sample \cite{hawai99}.
The most striking effect is the strong (almost sixfold)
change in resistivity $\rho(T)$ with temperature in the \lq\lq metallic\rq\rq\/ regime
of $d\rho/dT>0$. It can be shown that
the phonon scattering rate in Si-MOSFETs  is
negligibly low (less than 1\%) for the whole displayed
range of temperatures \cite{phonons}.

The data were obtained in the
temperature interval 0.3\,K to 45\,K and the temperature scale is
normalized to the Fermi energy $E_F$ in order to demonstrate
the  functional similarity of the strong drop in  $\rho(T)$
over the wide range of densities  $n$ from $\sim 1\times 10^{11}$ to
$35\times 10^{11}$\,cm$^{-2}$
(we use units of the temperature, K, for the energy;
 $E_F = 7.3 \times n$ [K$/(10^{11}$cm$^{-2})$].
 As temperature decreases,
the strong variation of  $\rho(T)$ seems to saturate
rather definitely at $T=(0.08-
0.1)E_F$. A closer look reveals, however,
that there is still (though much
weaker) temperature dependence even at lower temperatures. It is this dependence
which is the main subject of this paper.
\vspace{-0.2in}
\subsection{Phase diagram: From quantum to classical domains}
\vspace{-0.2in}
The range of high temperatures $T>E_F$ corresponds to a non-degenerate
carrier system; the corresponding
boundary
is depicted in Fig.~1 by
the vertical dash-dotted line (on the right).
In the opposite limit of much lower temperatures and
through the whole ``metallic'' range of densities, a
negative magnetoresistance is observed for $n-$SiMOS
samples in weak perpendicular
magnetic fields \cite{JETPL97,wl99}.
As is well-known, this effect is due to the suppression  of
quantum interference
\cite{lee&rama} and can be utilized to extract
the phase breaking time, $\tau_{\varphi}$, from the data.
In Refs.~\cite{wl99,wl2000},
$\tau_{\varphi}$ was found to fall off  with temperature as $T^{-1.5}$
whereas the transport mean free time, $\tau$,
 (as is seen from  $\rho(T)$ curves in Fig.~1) is
almost $T$-independent for
low temperatures. For that reason,
$\tau_{\varphi}$ drops faster than $\tau$  as $T$ increases
and at a certain temperature, $T_{\tau_\varphi =\tau}$,
the two times become equal. Evidently,
$T_{\tau_\varphi =\tau}$ defines a boundary
between the quantum-interference ($T<T_{\tau_\varphi =\tau}$) and
semi-classical ($T> T_{\tau_\varphi =\tau}$)
domains. This boundary is depicted in Fig.~1 by the dashed line
``$\tau_{\varphi}=\tau$''.
The domain ``SCl'' in Fig.~1  corresponds to the
semi-classical physics where quantum-interference effects are irrelevant.
{\em We emphasize that
the range of temperatures where the resistivity exhibits a
strong ``metallic''-like
$T$-dependence, belongs  almost entirely to the semi-classical domain
and is therefore a semi-classical effect}
(though not necessarily of a single-particle origin--it can be,
 e.g., a Fermi-liquid effect, such as T-dependent screening
\cite{screening}).
\begin{figure}
\begin{center}
\resizebox{3in}{3.8in}{\includegraphics{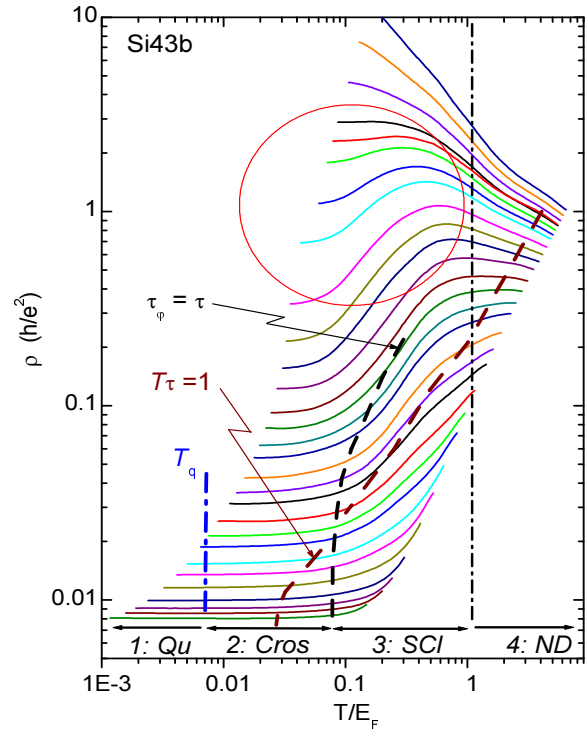}}
\begin{minipage}{3.2in}
\vspace{0.1in}
\caption{Resistivity
for sample Si43b as a function of the temperature normalized to
$E_F$. Densities, from top to bottom: 0.806, 0.8507, 0.8955, 0.94,
0.954, 0.985, 1.03, 1.075, 1.16, 1.25, 1.34, 1.49, 1.67, 1.85,
2.07, 2.30, 2.75, 3.19, 3.64, 4.54, 5.43, 6.33, 8.12, 9.91, 12.6,
17.1, 21.6, 26.0, 35.0
(in units of $10^{11}$\,cm$^{-2}$). Data are reproduced
from Ref.~\protect\cite{hawai99}, empiric
boundaries
(dashed and dash-dotted lines)
are depicted schematically  according
to the experimental results of
Refs.~\protect\cite{wl2000,gmax}.
Domain labels: {\em 1: Qu} - quantum,
{\em 3: SCl} - semi-classical, {\em 4: ND} - non-degenerate,
{\em 2: Cros}
- crossover between the semi-classical and quantum domains.
The meaning of other boundaries
is explained in the text.}
\label{fig1}
\end{minipage}
\end{center}
\end{figure}
\vspace{-0.2in}
Another important crossover temperature is
$T_{T=\hbar/\tau}$, at which
$T=\hbar/\tau$. For $T<T_{T=\hbar/\tau}$ interaction
corrections to the conductivity \cite{lee&rama,aa}
are expected to set in. The corresponding boundary
is denoted by ``T$\tau$=1'' in Fig.~\ref{fig1}.
Again, a large fraction of the $\{T,n\}$ domain
of the rapid, metallic-like $\rho(T)$-dependence
is to the right of the ``T$\tau$=1'' line, where
the quantum interaction effects are irrelevant as well.

One might expect
the quantum corrections to the conductivity to set in as $T$
decreases below the ``$\tau_{\varphi}=\tau$'' and
``T$\tau$=1'' lines.
In fact, the actual temperature at which the quantum correction becomes
observable may be significantly lower than the corresponding
crossover temperatures. For example,
in order
for the weak-localization correction to be larger than, e.g., 1\% of the total
resistivity, the ratio $\tau_{\varphi}/\tau$ has to be larger than
%$\exp(\pi 0.01\rho e^2/h)$,
$\exp[0.01\pi h/(e^2\rho )]$,
which is much larger than one for $\rho \lesssim \pi h/100e^2$.
%In order this inequality to be fulfilled in the region of a
%typical metallic resistivity $\rho \sim 0.01(h/e^2)$ (see Fig.~1),
%$\tau_{\varphi}/\tau$ must be bigger than $\exp(\pi)$. Taking into
%account that experimentally \cite{wl99,wl2000},  $\tau_{\varphi} \propto T^{-1} \div T^{-1.5}$,
%we obtain the estimate
%$T_q/T_{\tau_{\varphi}=\tau} \sim 8 \div 20$.
 Note
that even the latter estimate works only if the semi-classical
(Drude) resistivity does not depend on temperature. The
metallic-like temperature dependence of the Drude resistivity
pushes the onset of the  quantum-interference effects down to even
lower temperatures.

These
estimates
are in accord with an empiric finding
\cite{JETPLlnT,gmax} that the logarithmic quantum
corrections (with $d\rho/dT <0$) become well pronounced only starting
at $T/E_F < 0.007$ (or $T/T_{\tau_{\varphi}=\tau} \lesssim 10$); this
boundary is depicted by the left vertical line
``$T_q$''. It will be shown later in this paper that this empirical
boundary can be understood within the weak-localization theory for
a given measured $\rho(T)$-temperature-dependence in the semi-classical
region. In the crossover range, $ T_q<T<T_{\tau_\varphi =\tau}$,
as will be shown below, the ``localizing'' quantum correction to the
conductivity are masked  by a semi-classical $T$-dependence (with
$d\rho/dT >0$).

As density decreases and $\rho$ increases approaching the
insulating state, the negative magnetoresistance peak becomes
broader \cite{JETPL97,wl99,wl2000} which makes
identification of the quantum/semi-classical boundary
progressively more difficult.
The corresponding region in the vicinity of the MIT is
encircled in Fig.~\ref{fig1}.
No
quantum corrections to the conductivity have been quantified
experimentally in this region so far. On the other hand,
the theoretical perturbative results for quantum corrections to conductivity
are not supposed to be valid
in this range of $r_s \sim 8-10$ (here $r_s= E_c/E_F$ is the
ratio of the Coulomb to Fermi energy) and $\rho \simeq h/e^2$.
\vspace{-0.2in}
\section{Data analysis in the quantum and semi-classical domains}
\vspace{-0.1in}
\subsection{Experimental details}
\vspace{-0.1in}
In the rest of this paper,  we will use
the experimental data on $\rho(T)$ obtained on four Si-MOS
samples,
whose
relevant parameters are listed in Table 1.
\begin{minipage}{3.2in}
\begin{table}
\caption{Sample parameters.
$\mu_{peak}$ is the peak mobility at $T =0.3$\,K.
$n_c$ and $\rho_c$ are
the critical density (in units of $10^{11}$\,cm$^{-2}$) and critical
resistivity (in units of $h/e^2$), respectively.}
\begin{tabular}{|c|c|c|c|}
sample &
$\mu_{\rm peak} (\mbox{m}^2/\mbox{Vs})$ &
$n_{\rm c}$ &
$\rho_{\rm c}$  \\
\hline
Si-22/9.5 & 3.3 & 0.83 & 2 \\
Si-15a  & 3.2 & 0.82 & 2.5 \\
Si-43b  & 1.96 & 1.4 & 0.7 \\
Si-4/32 & 0.9 &  2.0 & 0.58 \\
\end{tabular}
\end{table}
\end{minipage}
\vspace{-0.3in}

\subsection{Localizing upturn in $\rho(T)$: weak-localization theory}
\vspace{-0.2in}
Within the conventional theory of disordered Fermi liquids
\cite{lee&rama,aa} which is applicable in the ``metallic''
domain ($\rho<\pi h/e^2$), weak localization and interaction
should eventually lead to the ``localizing'' ($d\rho/dT<0$)
$T$-dependence of the resistivity for low enough temperatures.
The competition between these two effects and the one(s) responsible
for the ``metallic''  ($d\rho/dT>0$) dependence at higher
temperatures should result in a resistivity minimum.
It is (partially) the lack of  observation of
of such a minimum
near the ``transition'' in Si MOSFET's
 as so far prevented an identification of the encircled region in
Fig.~1.
However, at higher densities the minimum
does occur \cite{JETPLlnT,gmax} within the
accessible temperature range. Also, the minimum
is observed near the transition in p-GaAs \cite{simmons_9910368}.
It is thus of crucial importance to understand
what are the predictions of the conventional
theory with respect to the position of the minimum
and its dependence on the carrier density and
other parameters.

In this regard, three of the authors (BLA, DLM, and VMP)
have recently put forward the following argument \cite{amp}.
Let us assume,
in accord with experimental data \cite{RvsT}, that
\begin{enumerate}
\item[i)]
the Drude resistivity is of the form
\beq
\rd(T)=\rho_1+\rho_0(T),
\eeq
where the temperature-dependent part, $\rho_0(T)$, is metallic-like
($d\rho_0/dT>0$) and much smaller than the $T$-independent one, $\rho_1$;
the latter assumption corresponds to the crossover domain ``2'' in
Fig.~\ref{fig1};
\item[ii)] the phase breaking time, $\tau_{\varphi}$, scales with $T$
as \cite{wl99,wl2000}:
\beq
\tau_{\varphi}(T)\propto T^{-p};
\label{tauph}
\eeq
\item[iii)] both the Drude and observable resistivities are small
compared to $h/e^2$, so that the weak localization (WL) theory is
applicable.
\end{enumerate}
It then follows immediately from the familiar
expression for the WL correction  in 2D \cite{lee&rama}
\beq
1/\rho=1/\rd(T)-\left(e^2/\pi h\right)\ln
\left[\tau_{\varphi}(T)/\tau(T)\right],
\label{weakloc}
\eeq
that under these conditions
the resistivity has a minimum at a temperature
determined from the following equation \cite{remark1}:
\begin{equation}
\tmin=\frac{e^2 p}{\pi h}\rho_{1}^{2}
\left(\frac{d\rho _{0}}{dT}\Big|_{T=\tmin}\right)^{-1}.
\label{tmin}
\end{equation}
(When differentiating Eq.~(\ref{weakloc}) with respect to $T$, we
neglect
the derivative $d\tau/dT$, whose contribution to (\ref{tmin}) is of the
order
of $\rd e^2/h$, which is much smaller than one
 within the WL theory and thus negligible.)
Interaction leads to another logarithmic term
in Eq.~(\ref{weakloc}) with a prefactor consisting of two
parts: a universal (singlet channel) and non-universal one
(triplet channel) \cite{aa}. This amounts to a change
in Eq.(\ref{tmin}): $p\to p^*$. Although $p^*$  may itself depend
on $r_s$ (via the corresponding dependence of
 the triplet channel contribution), we disregard this
possibility in the present analysis and view $p^*$
as a phenomenological parameter not necessarily equal
to $p$.

In the context of experiments on the \lq\lq metal-insulator transition
in 2D \rq\rq\/, the authors of Ref.~\cite{amp}
further observed that if  $d\rho _{0}/dT$ is
a {\it decreasing} function of the carrier density, $n$, then $\tmin$
{\it increases} with $n$. One arrives thus at a somewhat counter-intuitive
conclusion: the deeper one goes into the
\lq\lq metallic\rq\rq\/ regime, i.e., into the region of {\em larger} densities
and smaller resistivities,
the {\em higher} is the temperature at which
weak localization sets in.

This part of our paper has been criticized
in Ref. \cite{pp}.
It was pointed out there that the residual resistivity,
$\rho_1$,
depends on $n$ as well.
In particular,  at relatively {\it small} densities
[$n =(2.6-5.7)\times 10^{11}$cm$^{-2}$ \cite{JETPHall}]
$\rho_1$ {\it decreases} with increasing $n$.
This decrease can compensate the
decrease in $d\rho_0/dT$ with increasing
$n$
in the denominator of Eq.~(\ref{tmin}).
Extracting the $\rho_1(n)$- and $d\rho_0(n,T)/dT$-dependences from
the {\em small-$n$ data}
of Ref.~\cite{JETPHall} and using Eq.~(\ref{tmin}),
the author of Ref.~\cite{pp}
concluded that $\tmin$ {\it decreases} with increasing $n$ in the
density interval, specified above.
This would mean that the ``less metallic'' curves (closer to
the ``transition'') exhibit the weak-localization minimum
at higher temperatures than the ``more metallic'' ones
(further away from the ``transition''), which is not
consistent with the experiment.
On the basis of
this analysis, the author of Ref.~\cite{pp} claimed
 that the \lq\lq trap model\rq\rq\/,
suggested by two of us earlier \cite{am}, is \lq\lq
deconstructed\rq\rq\/.

We disagree with the conclusions of Ref.~\cite{pp}.
First of all, it should be pointed out
that the author
of Ref.~\cite{pp}
did not fully appreciate the fact
that the
derivation of Eq.~(\ref{tmin})
relies neither on the \lq\lq trap model\rq\rq\/ nor on any other
specific mechanism of the
temperature dependence of the
Drude resistivity. Thus if something
was \lq\lq deconstructed\rq\rq\/ in Ref.~\cite{pp},
it is the
applicability of the WL theory to the systems that
demonstrate
the \lq\lq metal-insulator transition in 2D \rq\rq\/.
Such a possibility can not be excluded a priori.
However, a mounting number of experiments, including those
presented in this paper, shows
that the \lq\lq anomalous metallic dependence
\rq\rq\/ is a semi-classical effect and
the \lq\lq anomalous metal \rq\rq\/demonstrates all the conventional
features of WL-behavior
\protect\cite{gmax,JETPLlnT,simmons_9910368,JETPHall,ensslin_9910228,hamilton_0003295},
including the localization upturn at lower temperatures.
This upturn was observed in Si MOSFETs at higher densities
($n \gtrsim 10-20n_c$, where $n_c$ is a suitably
defined \lq\lq critical density\rq\rq\/ of the transition)
\cite{JETPLlnT,gmax}, and in p-GaAs for
$n\approx n_c$ \cite{simmons_9910368}. On the phase diagram of
Fig.~\ref{fig1}, the upturn occurs in the vicinity of the empiric boundary $T_q$.

Assuming that the weak-localization theory
still describes the ``anomalous'' metallic state,
what density dependence of $\tmin$ does it predict?
To answer this question,
we performed a detailed analysis of the available experimental data.

It is shown in this paper that
no definite conclusion can be obtained via the procedure employed
in Ref.~\cite{pp}.
Extracting the
derivative, $d\rho/dT$, from the data is a an
ambiguous procedure: different
fitting functions, which approximate the data equally well, give
drastically
different results for $d\rho/dT$.
Disregarding this ambiguity for a moment,
for some fitting functions one obtains a {\it
non-monotonic} dependence of $\tmin$
on the carrier density. Indeed,
$\tmin$ first {\it decreases} with $n$
(in the density interval analyzed in Ref.~\cite{pp}),
but then it
goes through a minimum and
increases upon further increase in $n$.
Moreover,
given the discrepancy between results obtained
by using different fitting functions, one should not take seriously
any of the
conclusions obtained in this way,
including those of Ref.~\cite{pp}.
However, it is possible to carry out a much less
ambiguous analysis of the data.
As  it is  described below,
this analysis shows
that for Si MOSFET samples $\tmin$ {\it increases monotonically with $n$}.
Over a wide density range ($2.35n_c\leq n\leq 13 n_c$),
the calculated value $\tmin$ is smaller than the lowest temperature accessible
in the experiment in question ($T_{\rm acc} = 0.29$\,K).
The fact that no minimum in the resistivity
was observed for these densities receives thus a natural explanation:
the temperature was still too high to reach the minimum.
For larger densities, $\tmin$
is predicted to exceed $T_{\rm acc}$.
Indeed, for these densities the minimum in $\rho$ is observed.
Both the absolute value and the density-dependence of
the experimentally determined $\tmin$
is in a satisfactory agreement with the theory.
\vspace{-0.2in}

\subsection{Localizing upturn in resistivity: Experimental data}
\vspace{-0.2in}
A typical temperature dependence of the
resistivity was shown in Fig.~\ref{fig1}.
As the temperature decreases from
$\sim T_F$ to $\sim 0.1T_F$ (i.e., through the semi-classical
domain), the resistivity
decreases rapidly. A reasonably well description of the data
in this temperature range is given by the following
form \cite{pudJETPL97,gmax,akk}:
\beq
\rho(T) = \rho_1 +\rho_0\exp\left[T_0(n)/T)^p\right].
\eeq

\begin{figure}
\begin{center}
\vspace{-0.1in}
\resizebox{3.0in}{2.7in}{\includegraphics{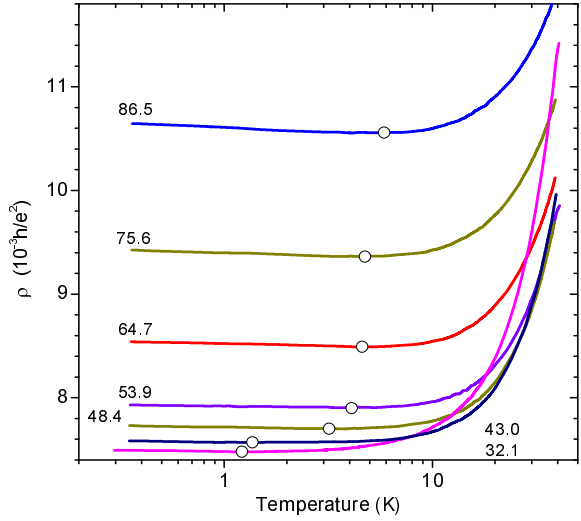}}
\begin{minipage}{3in}
\vspace{0.1in}
\caption{Temperature dependence of $\rho $
for sample Si15a.
Numbers at the curves indicate the density in units
of $10^{11}$cm$^{-2}$. Circles on the curves
mark positions of the minima.
%Re-plotted from Fig.~2 of
%Ref.~\protect\cite{gmax}.
Note that $dT_{\rm min}/dn >0$.}
\label{fig3}
\end{minipage}
\end{center}
\end{figure}

\begin{figure}
\begin{center}
\resizebox{3.3in}{2.9in}{\includegraphics{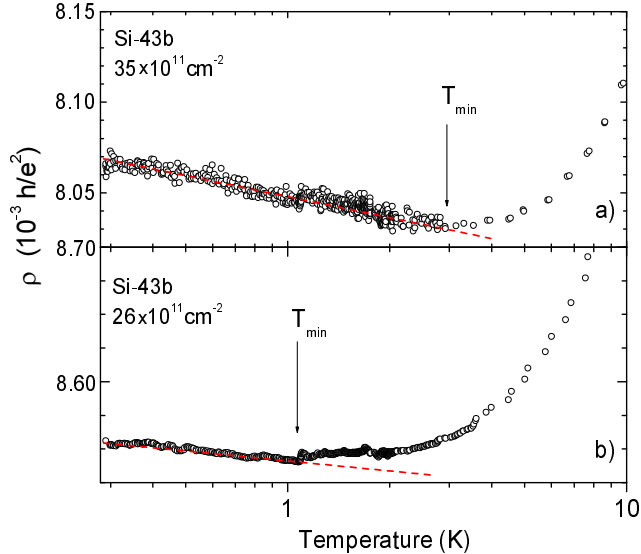}}
\begin{minipage}{3in}
\vspace{0.1in}
\caption{Expanded low-temperature part of the
$\rho(T)$-dependence for sample Si-43b at two different densities
in the range  $n>n^*$. Arrows mark positions of the resistivity
minima. Dashed lines: fit of the upward part of $\rho(T)$ with a
$\ln T$-dependence.}
\label{fig4}
\end{minipage}
\end{center}
\end{figure}

For lower temperatures
and higher densities, $n \geq n^*$
(where $n^*  \approx 20 \times 10^{11}$cm$^{-2}$ for this sample),
the strong drop in $\rho$
slows down, goes through a shallow minimum and
finally transforms into the
conventional  WL logarithmic
$T$-dependence. Fig.~\ref{fig3} shows the behavior of $\rho$
in the crossover regime between the domains of rapid variation
and WL upturn.
It is evident from Fig.~\ref{fig3}
that $\tmin$ {\em increases} with $n$.
%\vspace{-0.2in}
In order to demonstrate
the minimum in $\rho(T)$ in the range of \lq\lq high\rq\rq\/ densities
($n>n^*$) in more detail,
the crossover region for another sample,
Si-43b, is blown up in Fig.~\ref{fig4}.

%\vspace{-0.1in}

\begin{figure}
\begin{center}
\resizebox{3.0in}{3.0in}{\includegraphics{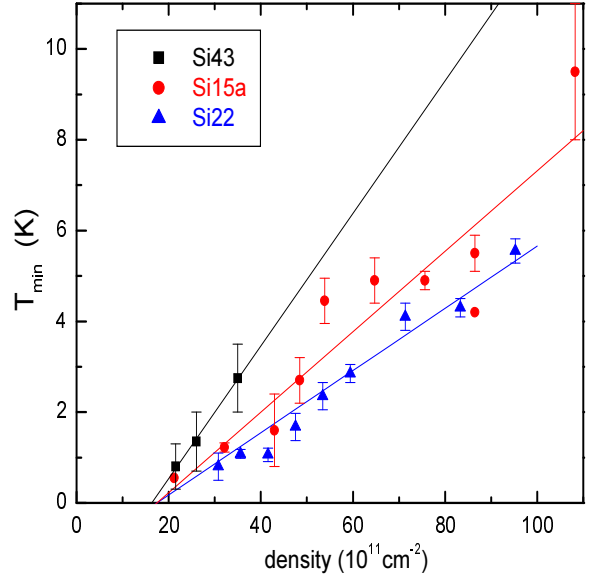}}
\begin{minipage}{3in}
\vspace{0.1in}
\caption{$T_{\rm min}$ as a function of the carrier
density for three samples. Solid lines are the guides for the eye.}
\label{fig5}
\end{minipage}
\end{center}
\end{figure}

We determined $T_{\rm min}$ from
our data
(in the  way demonstrated in Figs.~\ref{fig3}, \ref{fig4})
for those densities at which the minimum
is clearly pronounced.
 In Figure \ref{fig5}, the dependence of $T_{\rm min}$
on the carrier density is shown for three samples.
In a qualitative agreement with Ref. \cite{amp},
$T_{\rm min}$ decreases
with the density down to
$n=n^* \sim (20-25)\times 10^{11}$cm$^{-2}$.
For lower densities, $n<n^*$, the minimum in $\rho(T)$ is
probably below the lowest
accessible temperature of the experiment
%\cite{pudalovT/TF},
$T_{\rm acc}=0.27-0.3$\,K, and thus unobservable.
On the phase diagram Fig.~\ref{fig1}, this  occurs when
$T_q$ becomes equal  to  $T_{\rm acc}=0.3$\,K.

\vspace{0.1in}
%\vspace{-0.1in}

\begin{figure}
\begin{center}
\resizebox{3.1in}{3.6in}{\includegraphics{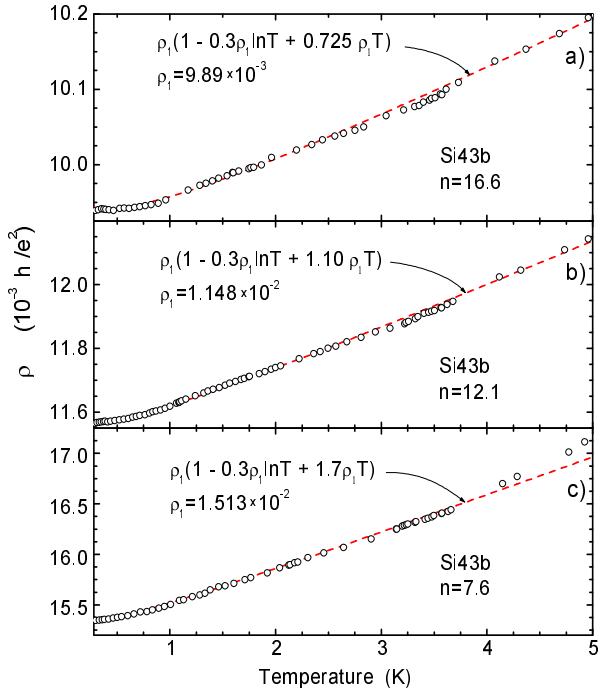}}
\begin{minipage}{3.2in}
\vspace{0.1in}
\caption{Expanded low-temperature part of the
temperature dependence of the resistivity for sample Si-43b at
three different densities in the {\em ``low density'' range} ($n <
n^*$). Dashed lines show the best fit of the experimental
$T$-dependences with Eqs.~(\ref{decomp},\ref{lint}).}
\label{fig6}
\end{minipage}
\end{center}
\end{figure}

For densities lower than $n^*$,
$\rho (T)$ {\em decreases with $T$} down to the lowest
accessible temperature (Fig.~\ref{fig6}).
This behavior was attributed earlier
to a {\em delocalizing logarithmic
temperature dependence} \cite{JETPLlnT}.
Similar results were also reported
in Ref. \cite{simmons_pGaAs,pepper99} for $p-$GaAs/AlGaAs.
However, a more detailed analysis of the data for Si-MOSFET samples
enabled us to decompose the
apparent ``metallic''\/ $T$-dependence of the conductivity, observed
in the range $T=(0.07-0.02)\times E_{F}$
into two contributions, namely,
\begin{equation}
1/\rho (T)=1/\rho _{{\rm lin}}(T)+B\ln T,
\label{decomp}
\end{equation}
where
\begin{equation}
%\rho _{{\rm lin}}(T)=\rho _{1}(n)+A(n)\frac{T}{T_{F}};\,A>0.
\rho _{{\rm lin}}(T)=\rho _1(n)\left[1 + A(n)\frac{T}{T_{F}}\right];\,A>0.
\label{lint}
\end{equation}

The first term in Eq.~(\ref{decomp}) dominates
in the crossover region,  $T \gtrsim (0.007-0.07)T_F$
in Fig.~\ref{fig1}, whereas
the second one corresponds to the conventional weak
localization correction (see Eq.~(\ref{weakloc})) and becomes more important
at lower temperatures ($T<T_q$ in the diagram of Fig.~\ref{fig1}).
\vspace{-0.1in}

\begin{figure}
\begin{center}
\resizebox{3.1in}{2.4in}{\includegraphics{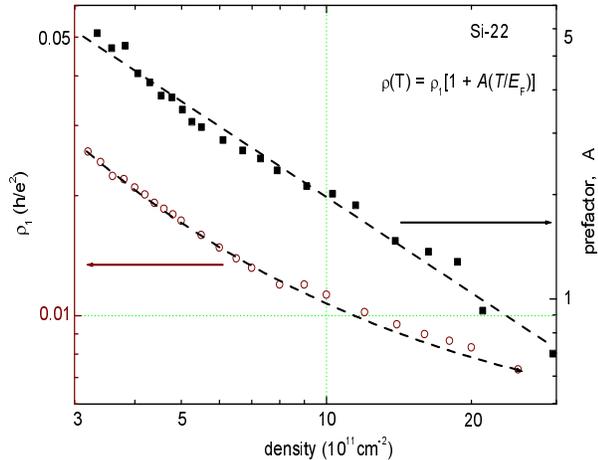}}
\begin{minipage}{3.2in}
\vspace{0.1in}
\caption{Density dependence of parameters $\rho_1$
and $A$ specifying the linear temperature dependence of the resistivity
measured for sample Si22
(see Eq.~(\ref{lint})). Dashed lines are guide for the eye.}
\label{fig7}
\end{minipage}
\end{center}
\end{figure}

Figure~\ref{fig6} shows the data for $\rho (T)$  for three different
densities, fitted with Eqs.~(\ref{decomp},\ref{lint}). The
density dependences
of fitting parameters $\rho _{1}$ and $A$
are shown in Fig.~\ref{fig7}. Coefficient $B$ turns out to be
almost independent of the density:
for $n =(5 - 30)\times 10^{11}$\,cm$^{-2}$
it equals to
$0.3\pm 0.1$.

%\vspace{0.1in}
\vspace{-0.1in}

There are little doubts
that the observed temperature
dependence of the resistivity is
indeed a semi-classical, metallic-like
behavior which masks the localization effect at not too large
densities and not too small temperatures,
rather than an indication of a novel
\lq\lq antilocalizing\rq\rq\/ behavior.
Similar conclusions regarding $p-$type GaAs/AlGaAs
and SiGe quantum wells were made in
Refs.~\cite{simmons_9910368,ensslin_0004312}.

We note  that the \lq\lq metallic\rq\rq\/
linear $T$-dependence of the resistivity is seen  best
of all in most disordered samples (see e.g., Fig.~1c in
Ref.~\cite{mauterndorf} and
Fig.~5 in  Ref.~\cite{akk}), in which the exponential
$\rho(T)$-dependence is much weaker
and does not mask the linear one.
This  linear $T$-dependence
is similar to that observed earlier \cite{wheeler,dolgojetp,plews}.
At first sight, it also seems to be consistent
 with the one predicted theoretically
for  the $T$-dependence of  screening
\cite{screening}.
In fact, whereas for high mobility samples
there is a rough similarity between the measured and predicted
density dependences of the prefactor $A(n)$,
for lower mobility samples  we found an essential
disagreement even on a qualitative level.
A much more detailed theoretical
and experimental work is required
to establish the precise nature
of the linear  $\rho(T)$-dependence. This information, however, is not needed
for our present purpose of translating the {\em measured} $\rho(T)$
dependence into  the position of the minimum
via Eq.~(\ref{tmin}).

\section{Comparison of experiment to  theory}
\subsection{How it is not to be done}
\begin{figure}
\begin{center}
\resizebox{3.2in}{2.8in}{\includegraphics{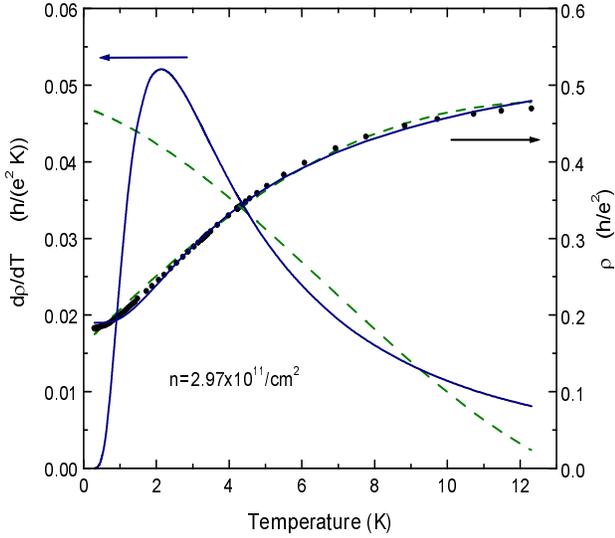}}
\begin{minipage}{3.2in}
\vspace{0.1in}
\caption{Right axis: experimental data for $\rho(T)$ (circles) for
$n=2.79\times 10^{11}$cm$^{-2}$ (sample Si4/32).
Solid and dashed lines: fits with Eq.~(\ref{exp})
and Eq.~(\ref{poly}), respectively.  Left axis: temperature derivatives of the fitting
functions.}
\label{fig8}
\end{minipage}
\end{center}
\end{figure}

\vspace{-0.2in}
We now turn to the analysis of the data in terms of the WL theory.
The evaluation of $\tmin$ from Eq.~(\ref{tmin})
appears to be rather
straightforward:
one fits the measured $\rho(T)$ with a suitable
function, finds the derivative of this function and then solves
transcendental
equation (\ref{tmin}) for $\tmin$.
This is what was done in Ref.~\cite{pp}.
Here we repeat
this procedure for a much wider range of densities
and for two different choices of the fitting function.
%As an example of the ambiguity one runs into on this way, we show
In Fig.~\ref{fig8} we demonstrate
an example of the ambiguity, which is inevitable in such
a procedure. The same
data for $\rho(T)$ (sample Si4/32,
$n = 2.97\times 10^{11}$cm$^{-2}$) is fitted with two functions:
exponential (solid)
\beq
\rho_{\rm exp}(T) =(h/e^2)\left\{0.19+0.4089\exp(-4.243/T[{\rm K}])\right\}
\label{exp}\eeq
and polynomial (dashed)
\beq
\rho_{\rm poly}(T)=(h/e^2)\sum_{n=0}^{4}C_n\left(T[{\rm K}]\right)^n
\label{poly}\eeq
where $C_0=1.610\times 10^{-1}, C_1=4.715\times 10^{-2},
C_2=-8.321\times 10^{-4}, C_3=-1.342\times 10^{-4}$, and $C_4=4.921\times 10^{-6}$.
As one can see from Fig.~\ref{fig8}, both function (\ref{exp}) and
(\ref{poly}) fit
the data equally well.
However, the temperature derivatives of fitting functions (\ref{exp}) and (\ref{poly})
differ drastically.
\vspace{-0.1in}

\begin{figure}
\begin{center}
%\vspace{-0.1in}
\resizebox{3.1in}{2.6in}{\includegraphics{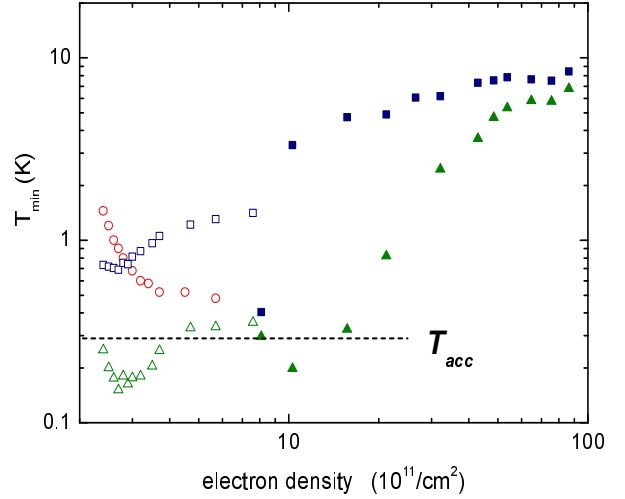}}
\begin{minipage}{3.2in}
\vspace{0.1in}
\caption{
Dependence of $\tmin$ obtained by the numerical solution
of Eq.~(\ref{tmin}) for $\rho_1$ and $\rho_0(T)$ from
the exponential [squares, Eq.~(\ref{exp})] and polynomial
[triangles, Eq.~(\ref{poly})] fit of the data.
Open symbols: sample Si22/9.5, solid symbols: sample Si15.
Horizontal line: lowest accessible temperature $T_{\rm acc}$.
Circles: $\tmin(n)$ reproduced from Ref.~\protect\cite{pp}.}
\label{fig9}
\end{minipage}
\end{center}
\end{figure}

The calculated values of $\tmin$ for both fitting functions (\ref{exp})
and (\ref{poly})  are shown in
Fig.~\ref{fig9} along with the results of Ref.~\cite{pp} (circles).
We have chosen $p=1$ for the coefficient in Eq.~(\ref{tmin}).
This is roughly consistent with the measured $T$-dependence of $\tau_{\varphi}$
for these samples. Triangles (squares) depict the results for the polynomial
(exponential) fit. Solid (open) symbols correspond to  sample Si15a (Si22/9.5).
Not surprisingly, the mentioned difference
in derivatives of fitting functions leads to
different values of $\tmin$,
both of which differ also from the result of Ref.~\onlinecite{pp}.
The tendency of $\tmin$ to decrease with increasing $n$,
which  was emphasized so strongly in Ref.~\onlinecite{pp},
manifests itself in our calculations as well.
{\it However, it occurs only at small densities}.
As the density increases
further, $\tmin$ goes through a minimum
and then starts to increase
with $n$.
We are not aware of what kind of fitting procedure
was used in Ref.~\cite{pp},
and how the ambiguity demonstrated above was avoided.
It is also not quite clear for us why the analysis of
Ref.~\cite{pp}
was restricted to a  rather narrow density range,
whereas the data over a much wider density  range
had been already available.
We are certain though
that
no definite statement can be
made on the basis of the results depicted in Fig.~\ref{fig8}.
\subsection{How it can be done}
\begin{figure}
\begin{center}
\resizebox{3.in}{2.2in}{\includegraphics{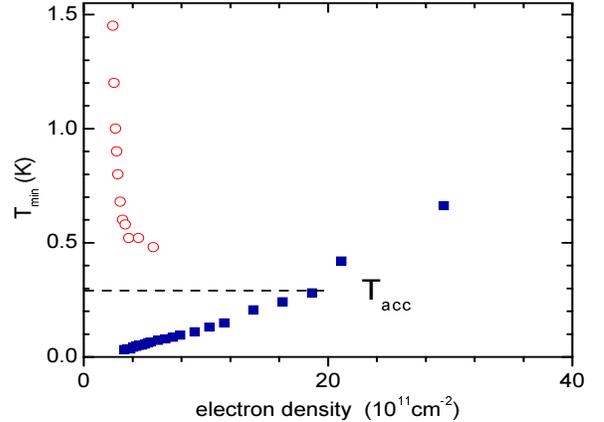}}
\begin{minipage}{3.2in}
\vspace{0.1in}
\caption{Full squares: $\tmin$ vs $n$ obtained by fitting  the
data for the sample Si22/9.5 with
Eqs.~(\ref{decomp},\ref{lint}). Note that in contrast
to Fig.~\ref{fig9}, $\tmin$ {\em increases} monotonically with $n$.
The absolute values of $\tmin$  are significantly smaller
than those  presented in Ref.~\protect\cite{pp}.
Horizontal line:  the lowest accessible temperature $T_{\rm acc}$.
For $n<20\times 10^{11}$\,cm$^{-2}$, our calculation gives
$\tmin<T_{\rm acc}$, which is consistent with the
direct measurements of $T_{min}$ shown in
Fig.~\protect\ref{fig5}.
$\tmin$ obtained in Ref.~\protect\cite{pp} (empty circles)
is larger than $T_{\rm acc}$,
which contradicts the experiment.}
\label{fig10}
\end{minipage}
\end{center}
\end{figure}

\vspace{-0.2in}
The separation of the measured dependence into two contributions
[cf. Eq.~(\ref{lint})]
allows one
to minimize the ambiguity of the fitting procedure.
The dependence of $\tmin$ on $n$, obtained from
Eq.~(\ref{tmin})
is shown in Fig.~\ref{fig10},
where we used the experimentally
determined $\rd(T)=\rho_{\rm lin}$
[Eq.~(\ref{lint})] data for the sample Si22/9.5.
As one can see,
$\tmin$ increases monotonically
with $n$ and crosses the lowest accessible temperature at $n\approx
20\times 10^{11}$cm$^{-2}$.
This is precisely the density $n^*$ above
which the minimum was actually observed (see Fig.~\ref{fig5}).
For $n\lesssim 6\times 10^{11}$cm$^{-2}$ the calculated value of $\tmin$
is below $100$\,mK.
For the lowest density, at which the fit with
Eqs.~(\ref{decomp},\ref{lint}) still
works reasonably well ($n=2.35n_c$), the calculated value
of $\tmin$ is $30$\,mK. The experimental difficulties arising from
electron heating at such low temperatures have been
discussed in detail in Ref.~\oncite{akk}.
%%%%%%%%%%%%%%%%%%%
\vspace{-0.2in}
\vspace{-0.1in}

The $\tmin(n)$ -dependence displayed in  Fig.~\ref{fig10}
is by no means universal.
It is determined by a particular form of the dependence
of the Drude resistivity on $T$, which varies from
system to system and even from sample to sample within
one system class.
A careful analysis of the
experimental
$\rho(T)$-dependence is
needed to extrapolate the
position of $\tmin$.
In light of this,
it can be readily understood
why some
p-GaAs samples \cite{simmons_9910368} exhibit a  resistivity  minimum
already for $n\sim n_c$,
and the temperature of the minimum
{\it decreases} with increasing $n$.

\section{Summary}
To summarize, we analyzed the available experimental data
for the resistivity of high mobility Si-inversion layers,
over a wide range of temperatures and densities.
Using recent experimental data on the
temperature dependence of the quantum coherence time \cite{wl2000}
we classified the domains of temperatures and densities where
quantum interference, classical and semiclassical effects
govern the conduction. We clarified experimentally
the various temperature
dependences dominating over these domains and analyzed them theoretically.
We found, in particular,
(i) the anomalous strong metallic-like temperature
dependence (with $d\rho(T)/dT>0$) belongs almost
entirely to the `high temperature' semiclassical domain $ E_F> T> \sim 0.1E_F$,
(ii) over the crossover domain $\sim 0.1E_F >T > T_q$,
an approximately linear metallic-like dependence
(most likely of the semiclassical origin), $\delta \rho(T) \propto
T/E_F$, dominates,
and (iii) over the quantum domain $ T < T_q$, the logarithmic $T-$dependence
(with $d\rho/dT < 0$) prevails.
Through a limited temperature (crossover) range,
the interplay of the two latter dependences
mimics the metallic-like behavior.
As the temperature is decreased further,
the  resistivity passes through a minimum and turns upward.
Using a phenomenological theory
we described the interplay of the two above effects  and made a
prediction on the temperature $T_{\rm min}$ where the resistivity shows an upturn.
We found a reasonably good
agreement between the experimentally determined
positions of these resistivity minima vs carrier density and
the calculated ones. In order to calculate
reliably the position of these minima, it is necessary to know precisely not
only  $\rho(T)$ in the low-temperature (crossover) region, but  also its
temperature derivative  $d\rho(T)/dT$;  the latter one
requires  highly precise data and their careful analysis.
Finally, we conclude that the existing data on the temperature
dependence of the resistivity of high mobility Si inversion layers
over the range of $\rho \ll h/e^2$ can  be successfully described
in terms of a temperature dependent semiclassical Drude
resistivity in combination with conventional weak localization
theory.

\acknowledgements
The work at Princeton University was supported
by ARO MURI DAAG55-98-1-0270. The work at Johannes Kepler
University of Linz was supported by FWF Austria
(P13439). D.\ L.\ M. acknowledges the
financial support from NSF DMR-970338 ``Mesoscopic Interacting Systems''
and the Research Corporation
Innovation Award (RI0082). D.\ L.\ M.\ and V.\ M.\ P.
are grateful for the partial support from
NSF DMR-0077825 ``Anomalous Metal in Two Dimensions''.
V.\ M.\ P.\ also acknowledges the support
from INTAS, RFBR, NATO, Programs ``Physics of solid state
nanostructures'', ``Statistical physics'' and ``Integration''.

\end{multicols}

\end{document}